# Snapshot 3D tracking of insulin granules in live cells

Xiaolei Wang*[a], Xiang Huang[b], Itay Gdor [a], Matthew Daddysman [a], Hannah Yi [c], Alan Selewa[d], Theresa Haunold [c], Mark Hereld[b] and Norbert F. Scherer[a,c]

[a] James Franck Institute, University of Chicago, 929 E. 57th St., Chicago, IL, USA 60637
[b] Mathematics and Computer Science Division, Argonne National Laboratory, 9700 S. Cass Avenue, Lemont, IL USA 60439
[c] Department of Chemistry, University of Chicago, 5801 S. Ellis Ave., Chicago, IL USA 60637
[d] Biophysical Sciences, University of Chicago, 924 E. 57th Street, Chicago, IL USA 60637

**ABSTRACT**

Rapid and accurate volumetric imaging remains a challenge, yet has the potential to enhance understanding of cell function. We developed and used a multifocal microscope (MFM) for 3D snapshot imaging to allow 3D tracking of insulin granules labeled with mCherry in MIN6 cells. MFM employs a special diffractive optical element (DOE) to simultaneously image multiple focal planes. This simultaneous acquisition of information determines the 3D location of single objects at a speed only limited by the array detector's frame rate. We validated the accuracy of MFM imaging/tracking with fluorescence beads; the 3D positions and trajectories of single fluorescence beads can be determined accurately over a wide range of spatial and temporal scales. The 3D positions and trajectories of single insulin granules in a 3.2um deep volume were determined with imaging processing that combines 3D decovolution, shift correction, and finally tracking using the Imaris software package. We find that the motion of the granules is super-diffusive, but less so in 3D than 2D for cells grown on coverslip surfaces, suggesting an anisotropy in the cytoskeleton (e.g. microtubules and action).

## 1. INTRODUCTION

Rapid 3-dimensional imaging is crucial in modern optical microscopy for understanding dynamics in and function of living cells. Three-dimensional objects and trafficking pathways of interesting targets (e.g. different proteins or vesicles) are not constrained to one plane[1, 2]. However, simultaneously obtaining all the information for accurate 3D imaging and particle tracking, termed snapshot 3D imaging, remains a challenge. In most recent imaging technologies, such as epi-fluorescence wide-field, spinning disk confocal and light sheet microscopy[3], only one focal plan is in focus at one time. Thus, sequential scanning is needed to obtain volumetric information. These modalities are, therefore, slow[4] and might miss important aspects of the dynamic events during scanning. Sequential scanning could be more susceptible to photobleaching for fluorescence probes due to long time required for scanning. Also, the requisite mechanical movements could generate systematic errors.

There are some emerging technologies to mitigate and potentially overcome these limitations, including lattice light sheet microscopy[5] and multifocal (plane) microscopy[6-9]. Variants of the latter are particularly promising for simultaneously obtaining all the 3D information in one snapshot. Multifocal plane microscopy was first demonstrated by Ober and co-workers[8] in a fluorescence microscope equipped with a beam splitter to image two focal planes on separate cameras, and then improved to image up to four focal planes[9]. This method has high light efficiency and a large field of view, but requires one camera per focal plane. Abrahamsson and co-workers developed an aberration-corrected multifocal microscope (MFM) with a diffractive multifocus grating, chromatic correction grating and multifaceted refractive prism[6]. MFM has a practical advantage of requiring just one camera, although that comes at the expense of non-trivial optical fabrication and a reduced field of view.

Quantitative analysis of insulin granule tracking in three dimensions in live cells is important for understanding intracellular transport, regulation of insulin secretion and diabetes[10, 11]. However, snapshot acquisition of 3D data with sufficient speed for accurate 3D tracking is the main challenge for dynamic analysis of insulin granules. MFM employs a special diffractive optical element (DOE) that behaves as a 2-dimensional diffraction grating and has a lens "distortion" in its design. This geometrical distortion compensates wavefront phase error from the out of focus plane to be corrected to a flat wavefront. Light from a point source at the nominal focal plane has a flat wavefront in Fourier space, while light from defocused (out-of-plane) sources has a parabolic (curved) wavefront. The diffraction and geometrical distortion of the DOE will cause the beams from the out-of-focus planes (i.e., those away from the primary image plane in Fig.1a) to become parallel; each focal plane corresponds to a different diffraction order of the DOE and the associated lens distortion causes them to simultaneously focus on camera using a single relay lens f2 (Fig. 1a). The MFM allows simultaneous imaging of many different focal planes[12] (e.g., 7, 9, 25 depending on the DOE design) for simultaneous acquisition of the information required to determine the 3D location of, in our case, single insulin granules at a speed limited by the frame rate of the array detector.

We demonstrate the utility of MFM for fast volumetric imaging of insulin granules labeled with mCherry in MIN6 cells[13, 14]. We designed the DOE focal shift to span a 3.2um deep volume. The 3D positions of single insulin granules and trajectories are determined through image processing that combines decovolution filtering with tracking using the Imaris software package. We also validated the accuracy of MFM imaging/tracking using immobilized fluorescence beads; the 3D positions and trajectories of single fluorescence beads can be determined accurately over a wide range of spatial and timescales. The processed MFM data allows quantifying the complex 3D dynamics of single insulin granules in MIN6 cells.

## 2. METHODS

### 2.1 Multifocal microscope setup

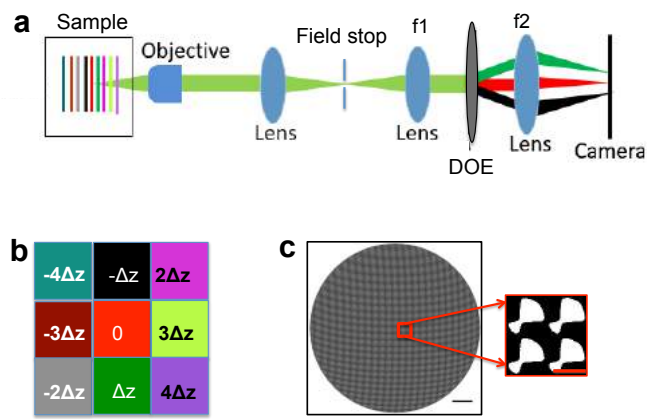

Figure 1. Multifocal microscope (MFM), focal-shifted tiles and the DOE optic. (a) The schematic of a MFM. Different colors represent different focal planes. f1 and f2 are relay lens with 200mm and 400mm focal length, respectively. The DOE is located at the second pupil plane, which the first relay lens generated. All nine focal planes are focused on the camera. (b) A schematic MFM image on the camera showing 9 titles imaging different focal planes. (c) The DOE we designed and a zoomed-in view that shows 4 unit cells of the DOE binary phase pattern. Scale bar (black) is 1mm, and scale bar (red) for four period of MFM is 50 microns.

A home-built multifocal microscope was developed based on Abrahamsson's design [6] as an extension to a commercial (Nikon Eclipse Ti) inverted microscope, as shown schematically in Figure 1a. The significant difference from that design is that we do not use chromatic correcting elements; i.e., the chromatic correction grating and multifaceted prism. The trade-off for the more simple (and less expensive) optical design is that additional image processing is required to account for the different the chromatic aberration in each focal plane tile. The microscope is housed in a temperature controlled and acoustically damped enclosure. A solid-state LED light source (Lumencor, Spectral X light engine) with highly stable output was coupled to the microscope with a liquid light guide. Excitation light was bandpass filtered (Semrock FF01-578/21-25) with maximum power of 515mW. Samples were mounted on a mechanical stage (Ludl

99S106-N2K-LE2, controller MAC5000). Fluorescence emission collected by the objective (Nikon 60x 1.27 NA CFI Plan Apo water immersion, MRD07650) was separated from the excitation light by a dichroic mirror (Semrock, FF596-Di01-25x36) and bandpass filter (Semrock, FF02-641/75-25). For MFM, a field stop was placed at the primary image plane, which crops the field of view to 2mm$^2$. The field stop consists of a circular 25.4 mm diameter by 1 mm thick fused silica substrate (Thorlabs WG41010) with a 200 nm Cr layer deposited on the entire surface except for a 2 mm square in the center. A 4-f lens system relays the image from the primary image plane onto the second image plane, where the camera (Andor iXon Ultral 888 EMCCD) is located. The first lens (f=200mm, Thorlabs AC508-200-A) matches the focal length of the tube lens (Nikon, f=200mm) and generates the secondary pupil (Fourier) plane for the MFM diffractive optical element (DOE). The second lens (f=400mm, Thorlabs AC508-400-A) focuses the fluorescence emission from the DOE onto the EMCCD detector. A second narrower bandpass filter (Semrock, FF01-620/14-25) is placed after the DOE to narrow the bandwidth of the emission light to mitigate chromatic aberration. The MFM imaging system simultaneously produces a focal stack of high-resolution 2D images on a single camera as shown in Fig. 1b. The 120x magnification gives a theoretical lateral pixel size of 108nm, which was confirmed by imaging a ruled slide. The lateral field of view in each MFM tile is 35.5 $\mu$m by 35.5 $\mu$m.

## 2.2 Diffractive optical element design and fabrication

The MFM DOE was designed with a period of 56 $\mu$m and focal shift $\delta z$ of 400nm (Fig. 1c) based on a procedure outlined by Abrahamsson et al. [6]. Our DOE is designed for a (fluorescence) wavelength of 610nm with emission light distributed evenly and efficiently (67%) among the nine tiles. We fabricated the DOE in-house on a 5 mm thick UV fused silica substrate (WG41050) that was cleaned with acetone, isopropyl alcohol and distilled water; then spin-coated with 1.5 $\mu$m thick layer of AZ1512 photoresist (Shipley). A laser writer (Heidelberg MLA150) exposed the desired pattern in the photoresist (405 nm laser at a dose of 150 mJ). The photoresist was developed in AZ-300 MIF developer (Integrated Micro Materials) for 20 s. Then the pattern was etched in a reactive ion etcher (RIE, Oxford Instruments). A depth of 680nm for the design wavelength of 610nm was etched by using gases Ar (25 standard cubic centimeters per minute (sccm) and CHF$_3$ (25 sccm)), and RF power of 200 W. The photoresist remaining after etching was stripped with acetone in an ultrasonic bath. The etching depth and surface roughness of the DOE were measured with contact profilometry. All fabrication steps were completed at the Prizker Nanofabrication Facility at the University of Chicago. The DOE binary pattern created with this procedure is shown in Fig. 1c.

## 2.3 Sample preparation

We devised stable CRISPR-Cas9 labeling of Insulin-2 in clonal MIN6 cells. The insulin-GFP guideRNA (gRNA) was designed using a previously characterized insulin-GFP plasmid, where the GFP is integrated immediately downstream and in-frame with the insulin C-peptide sequence. The insulin-mCherry construct was designed similarly by aligning the mCherry sequence to the original insulin-GFP construct. A CRISPR plasmid was generated using the Optimized CRISPR design tool to identify an appropriate cut site near the insulin-2 gene. The gRNA sequence was then incorporated into a plasmid that contained Cas9 protein. Mouse Insulinoma 6 (MIN6) cells were isolated, grown from single cells, and transduced at low passage with the CRISPR plasmids. The resulting fluorescent cells were isolated through three rounds of flow cytometry to generate a pure population of MIN6 cells expressing mCherry-Ins2.

The MIN6 cells expressing mCherry-Ins2 were cultured in high-glucose DMEM (Life Technologies 10569) supplemented with fetal bovine serum (10%, Life Technologies 26140) and penicillin-streptomycin (200 U/mL, Life Technologies 15140) under standard atmosphere conditions (5% CO$_2$, 37$^o$C). Cell cultures were trypsinized at 30-40% confluence and transferred to glass-bottom dishes (Mattek P35G-1.5-14-C) at 70-80% confluence one day before microscopy measurements. Before imaging, the sample was washed (1mL x 2), immersed in 1 mL of 28mM glucose in Krebs-Ringer buffer (pH of 7.4, 132 mM NaCl, 4.7 mM KCl, 10 mM HEPES, 2.5 mM CaCl$_2$, 1.2 mM MgSO$_4$, 1.2 mM KH$_2$PO$_4$, and 2 mM NaHCO$_3$), and allowed to incubate for 15 min.

## 2.4 MFM image reconstruction

To reconstruct a 3D image from the set of nine 2D multifocal images required arranging the nine focal plane images in order of the focal shift, aligning them laterally using the flood image, deconvoluting each image with the cognate measured point spread function using Huygens, and aligning them into a 3D stack with a transformation matrix approach[15] that is calibrated using 200 nm fluorescent beads (ThermoFisher, F8810) immobilized on a coverslip. A flood image was obtained with a fluorescent plastic slide (Chroma) with an excitation wavelength of 575nm, which gives the borders of the nine tiles using a threshold of the flood image that is used for cropping images of beads and granules.

Figure 2a is an MFM image of three 200nm fluorescent beads and 2b is the associated uncorrected image. Figure 2c shows side-on planarized projections of the 9 focal plane shifted images following coarse alignment based on the flood image. It is clear that there is an x, y shift between different focal planes and significant achromatic aberration. Elimination of these is crucial to obtaining accurate 3D imaging reconstruction. The point spread function (PSF) that allows removing the achromatic aberration with deconvolution was obtained with 100nm fluorescent beads (ThermoFisher, F8801). A focal series is measured by scanning the bead sample along the z axis with 50nm steps. In this z stack, the beads were successively focused in each tile. Nine 3D PSFs, one for each tile, were obtained with PSF distiller in Huygens (Scientific Volume Imaging, Hilversum, The Netherlands). Chromatic aberration and the out of focus background were removed for each focal plane (different tiles) by decovolution with these nine 3D PSFs using Huygens. The positions of maximum intensity of the three 200nm fluorescent beads in each tile (different focal planes) were used to determine the transformation matrix for fine alignment; i.e., by translation, rotation and scaling. Finally the nine deconvoluted focal planes were accurately superimposed with the transformation matrix (Fig.2d, Fig.2e and Fig.2f).

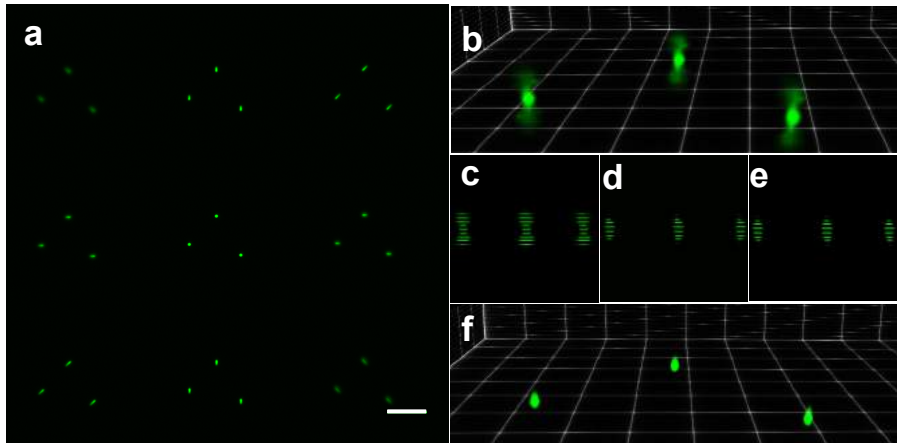

Figure 2. MFM image 3D reconstruction. (a) A single frame (image) of MFM data (9 tiles/focal planes) of three 200nm fluorescent beads. Scale bar is 10 $\mu$m. (b) 3D image of fluorescent beads just with cropping based on alignment using the flood image. (c), (d) and (e) are xz projections of raw data, data processed with deconvolution, and date processed with deconvolution and shift correction, respectively. (f) The final 3D image of fluorescent beads with deconvolution and shift correction.

## 3. IMAGING EXPERIMENTS

### 3.1 Immobilized fluorescent beads sample imaging for MFM verification

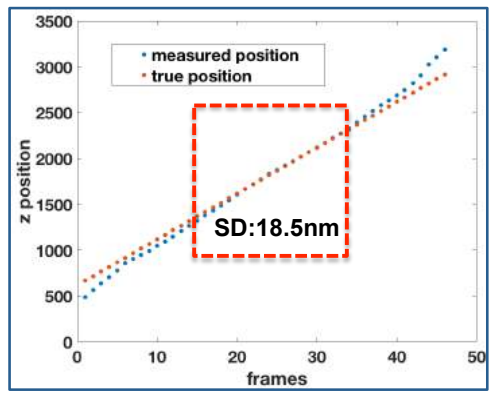

Figure 3. Determining the accuracy of MFM imaging and image reconstruction by scanning an immobilized 200nm fluorescent bead in the axial (z) direction. The red dotted line and blue dotted line respectively show the true position determined with the piezo nano-positioning and the measured position obtained by MFM. The standard deviation is

18.5nm in the z range defined by the red dash square (from frames14 to 34). A gradual systematic trend is also noticeable.

Estimating the accuracy of position determinations of single particles is a crucial task. Accuracy refers to the closeness to the true value, which can establish a potential systematic error or bias. Here we used an immobilized 200nm fluorescent bead to evaluate the accuracy of our MFM. To obtain images of fluorescent beads with different axial (z) positions, the sample stage was moved with a piezo nano-positioner (MadCity, Nano-Z200) in 50nm steps. The z position of the fluorescent beads was determined with MFM imaging. We obtained a trajectory of the fluorescent bead using 532mW illumination power, an exposure time of 0.5ms and EM gain of 30, which gave about 100,000 total photons per frame. MFM 3D image reconstruction was performed as described above and particle tracking was done in 3D using Imaris. The z location estimates for the bead over a range of value from 0.5 to 3.2$\mu$m illustrates the 50nm step movement of the piezo nanopositioner as shown in Fig. 3, where blue dots are the positions estimated from measurement and the red dots are the true positions. The bead is in focus in the center tile at frames 23. According to Fig.3, the difference is smallest when the bead is in-focus in the center tile, and becomes larger as the bead moves out of focus. When the bead is in focus in the center tile, it will get more photons in a specific 3D volume, which increases the precision of localization. The standard deviation of difference between measured position and true position is 18.5nm in the 1 $\mu$m z range indicated by the red dashed box. Fig.3 shows that MFM can correctly recover the z position values over the range of 2.7 $\mu$m.

### 3.2 Live MIN6 cell imaging

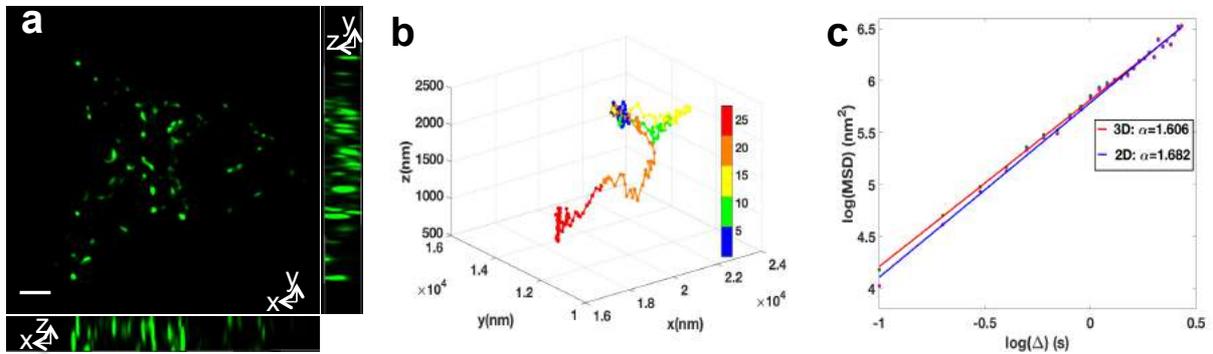

Figure 4. Single insulin granule tracking in an MIN6 cell measured by MFM. (a) A reconstructed 3D image of mcherry labeled insulin granules in a MIN6 cell (scale bar is 3 micro). The center tile is the xy projection. The two insets (at right side and bottom side) are yz and xz projections. (b) shows the trajectory of a single insulin granule. The time position of each tracked point is encoded by a different color (see the color bar). The unit of time is one second. (c) is a log-log plot of mean square displacement as a function of lag time $\Delta$ for 3D and 2D trajectory. The dots are measured results, and the solid lines are linear fitting.

Insulin is a hormone released by pancreatic islet β-cells[10, 11], in response to elevated blood-glucose levels to maintain glucose homeostasis. Failure of secretion is considered to be the main factor for type 2 diabetes. Insulin molecules are packed into granules to transport in cells. The stimulus-secretion process is biphasic– the rapid exocytosis that occurs within the initial 5-10 minutes of stimulation is described as the first phase and the subsequent release is referred to as second phase secretion. It's hypothesized that different "pools" of granules and membrane docking mechanisms play important roles in biphasic secretion[10]. The study of intracellular transport of insulin granules is therefore not only important to understand the fundamental principles of insulin secretion, but also to investigate diagnostics and therapeutics.

With MFM, we imaged 3D movement of insulin granules with one snapshot within a depth of 3.2 $\mu$m in the cell for 100s at 10 frames per second. We used CRISPR to tag mCherry to insulin, which is packaged into granules in MIN6 cell due to the high efficiency of labeling [16]. The height of monolayer grown MIN6 cells is around 6 or 7 $\mu$m. The layer that is within 2 $\mu$m of the substrate, which is adjusted to be in-focus in the center tile, gives the best ratio of signal to background. The green dots in fig.4a are a 3D image reconstruction of MIN6 cell. The two insets are xz and yz image projections clearly show different insulin granules have different z positions. We determine that the movement of insulin granules is 3-dimensional. Fig. 4b shows a 3D trajectory of a single insulin granule, which has a range of z displacement of around 2$\mu$m.

Analysis of single insulin granule trajectories provides insight into intracellular transport. The time averaged mean square displacement (MSD) is calculated for each single insulin granule trajectory as

$$\overline{\delta(\Delta,t)^2} = \frac{1}{t-\Delta}\int_0^{t-\Delta}[r(\tau+\Delta)-r(\tau)]^2\,d\tau \qquad (1)$$

where r is the position of single insulin granules, Δ is the lag time, and t is the total measurement time for each trajectory. Fig. 4c shows a log-log plot of the MSD of the trajectory of a single insulin granule shown in Fig.4b. As the lag time (Δ) increases, the range available for averaging decreases, so we use 10% of total measurement time of each trajectory to estimate the transport parameters. The scaling relation between MSD and lag time can be expressed as

$$\overline{\delta^2} = D\Delta^\alpha \qquad (2)$$

where D is the diffusion coefficient and the exponent α characterizes motion: α=1 for Brownian motion, α>1 for superdiffusive motion, and α<1 for subdiffusive motion. The linear fitting of log MSD vs. log lag time shown in Fig.4c indicates that the 3D motion of the insulin granule is superdiffusive, which means the motion of the insulin granule is directed. Since each tile of the MFM image is a 2D wide field image, we analyzed the 2D (xy) motion of the same insulin granule using the center tile of images. As shown in Fig. 4c, α is larger for 2D motion of the insulin granule in the xy plane than it is for 3D motion, while the diffusion coefficient is smaller. This result means that motion in the xy plane is different from motion along z. In particular, the z motion was found to be sub-diffusive. The quantitative analysis of single insulin granules in cells demonstrates the significance of obtaining 3-dimensional dynamical information for heterogeneous systems such as intracellular environments.

## 4. DISCUSSION

We developed a multifocal microscope (MFM) to image multiple focal planes simultaneously, which enabled us to track 3D motion in live cell environments. Our study demonstrates the feasibility of complementing MFM with image processing suitable for 3D intracellular tracking. For single particle tracking, the accuracy of 3D localization can be determined. We tested our MFM system and imaging processing with immobilized 200nm fluorescence beads showing that the accuracy of z position is better than 20nm when the bead (or granule) is within 1 μm of the focal plane imaged on the center tile of the 9-tile MFM image. We determined that the motion of single insulin granules in monolayer-grown MIN6 cells is superdiffusive, consistent with the understanding that insulin granules walk on the microtubules. Our finding that granule motion has a larger α value in 2D vs 3D suggests that z-motion is much more restricted than in planes parallel to the substrate that the cells are grown on. This may be due to an anisotropy of microtubule growth so that they tend to along parallel to the glass substrate within the cell causing restricted motion along z. The is, the structure of microtubules in monolayer-grown MIN6 cells is heterogenous.

Until the present report, our understanding of transport of insulin granules in cells came from 2D measurements using conventional (e.g. confocal) microscopy. However, the actual trafficking of insulin granules in cells is 3D, and we found the motion to be spatially anisotropic. This suggests that conventional microscopy is not suitable to study 3D dynamics due to the sequential imaging; when one plane of the cells is imaged, the information from other planes will be missed. Uncovering the complex 3D motion (of insulin granules) in cells requires novel optical microscopy such as MFM. Although our study points to the key advantage of MFM for camera limited imaging speed, photobleaching of fluorescent dyes and phototoxicity for living cell [17] still limits the photons collected per frame which limits 3D localization with high accuracy. Moreover, the large background for thick samples will become an issue because MFM is a wide-field image technique where each image contains out of focus information from other planes. We are addressing these issues in coming work.

## ACKNOWLEDGEMENTS

This work was supported by funding through the Biological Systems Science Division, Office of Biological and Environmental Research, Office of Science, U.S. Department of Energy, under Contract DE-AC02-06CH11357.